\documentclass[aps,prc,preprint,showpacs,showkeys,floatfix,nofootinbib,%
superscriptaddress]{revtex4}
\usepackage{graphicx}
\usepackage{multirow}
\usepackage{bm}

\newcommand{\ie}{{\textrm{i.e.}}}

\begin{document}
%%%%%%%%%%%%%%%%%%%%%%%%%%%%%%%%%%%%%%%%%%%%%%%%%%%%%%%%%%%%%%%%%%%%%%%

\title{Renormalization group analysis of the chiral pion production operator
for \boldmath$NN\to d\pi$}

\author{Satoshi X. Nakamura}\email{snakamura@triumf.ca}
\affiliation{Theory Group, TRIUMF,
4004 Wesbrook Mall, Vancouver, BC V6T 2A3, Canada}
\author{Anders G{\aa}rdestig}\email{anders@physics.sc.edu}
\affiliation{Department of Physics and Astronomy,
University of South Carolina, Columbia, SC 29208, U.S.A.}

\begin{abstract}
We are interested in  the consistency between the cutoff,
chiral symmetry, and the power counting.
For this purpose, we apply the Wilsonian renormalization group (RG)
to an operator and then decrease the Wilsonian cutoff.
As an example, we study the $s$-wave pion production 
operator for $NN\to d\pi$, derived in chiral perturbation theory.
We find that the renormalized part of the RG effective operator is accurately 
absorbed by chiral counter terms of higher order with natural coefficients.
Thus, the use of the (sharp) cutoff regularization does not require us to
introduce chiral-symmetry-breaking counter terms, at least
in the case of the $NN\to d\pi$ reaction.
\end{abstract}

\preprint{TRI-PP-07-28}

\maketitle

\section{Introduction}

Chiral symmetry and its spontaneous breaking are important low-energy
properties of QCD, and therefore it is natural to hope to incorporate them
into a description of a nuclear system.
A promising solution is to employ the chiral perturbation theory
($\chi$PT), which has been successfully used in nucleonic systems since 
the beginning of the 1990's.
The nuclear operators are derived from a chiral
Lagrangian following a counting rule.
The transition matrix element is evaluated by convoluting the operator
with nuclear wave functions.
In the evaluation, a cutoff is usually applied to the operator, suppressing
high momentum modes.
The question then arises whether this regulated operator is still consistent 
with chiral symmetry.
If not, we may need to introduce a chiral-symmetry-violating operator to
recover the symmetry.
Although it is conventionally supposed that effects of the high momentum
modes are captured by higher order chiral counter terms,
it is important to confirm this.
We consider a renormalization group (RG) analysis to be useful for this
purpose.

In this work, we apply the Wilsonian Renormalization group (WRG) equation
to the $s$-wave pion production in $NN\to d\pi$ as described by 
$\chi$PT\cite{lensky}.
We introduce a cutoff, which, when reduced using the WRG equation,
result in an operator that runs.
In the next step, we try to reproduce the RG low-momentum operator using
the starting operator plus chiral counter terms of higher order.
If the chiral symmetry is not violated by the cutoff, the renormalized
part of the operator should be accurately captured by the chiral counter
terms.

\section{Wilsonian RG equation for operator}
\label{sec_wrg}

At first, we discuss why we use the WRG equation to introduce and/or
reduce the cutoff.
This is based on a consistency with the construction of an effective
Lagrangian.
An effective Lagrangian can be obtained formally via a path integral 
formulation based on the Lagrangian of the underlying, more fundamental theory.
One then integrates out the high energy degrees of freedom.
When integrating out the high momentum states of the nucleon in the 
heavy-baryon $\chi$PT Lagrangian, we can also use the path integral.
This procedure is equivalent to solving the WRG equation derived below.
One of the present authors derived the WRG equation for the $NN$
interaction in this way in Ref.~\cite{rg1}.
The WRG equation for a transition operator ($\pi$ production operator in
our case) can also be derived in essentially the same way.

Here we outline a simpler derivation of the WRG equation following
Ref.~\cite{NA} in which a detailed account is given (Appendix A).
We start with a matrix element in which the transition operator is defined in 
a model space spanned by plane wave states of the two-nucleon system.  
The maximum magnitude of the relative momentum in this model space is
given by the cutoff $\Lambda$.
We differentiate the matrix element with respect to $\Lambda$ and
impose the renormalization condition that the matrix element is
invariant under cutoff changes.
This gives the WRG equation for the low-momentum transition operator.
(For explicit expressions, see Refs.~\cite{NA,NA2}.)
This differential equation is solved, and the solution in
integral form is actually the same as the effective operator in the
Bloch-Horowitz formalism.

\section{Results}\label{sec_result}
We start with the chiral NLO $s$-wave pion production operator for the
$NN\to d\pi$ reaction\cite{lensky}.
Using the integral form of the WRG equation,
we calculate the RG low-momentum operator for $\Lambda$ = 500~MeV,
using the CD-Bonn $NN$-potential.
We employ near threshold kinematics, \ie, $\eta$ = 0.1, 
where $\eta=q/m_\pi$ is the emitted pion 
momentum divided by the pion mass.
We use the chiral counter terms of higher order \{see Eq.~(7) of
Ref.~\cite{NA2} for expressions\} to
simulate the renormalized part of the low-momentum operator.

\begin{figure}
\begin{minipage}[t]{68mm}
\begin{center}
\includegraphics[width=65mm]{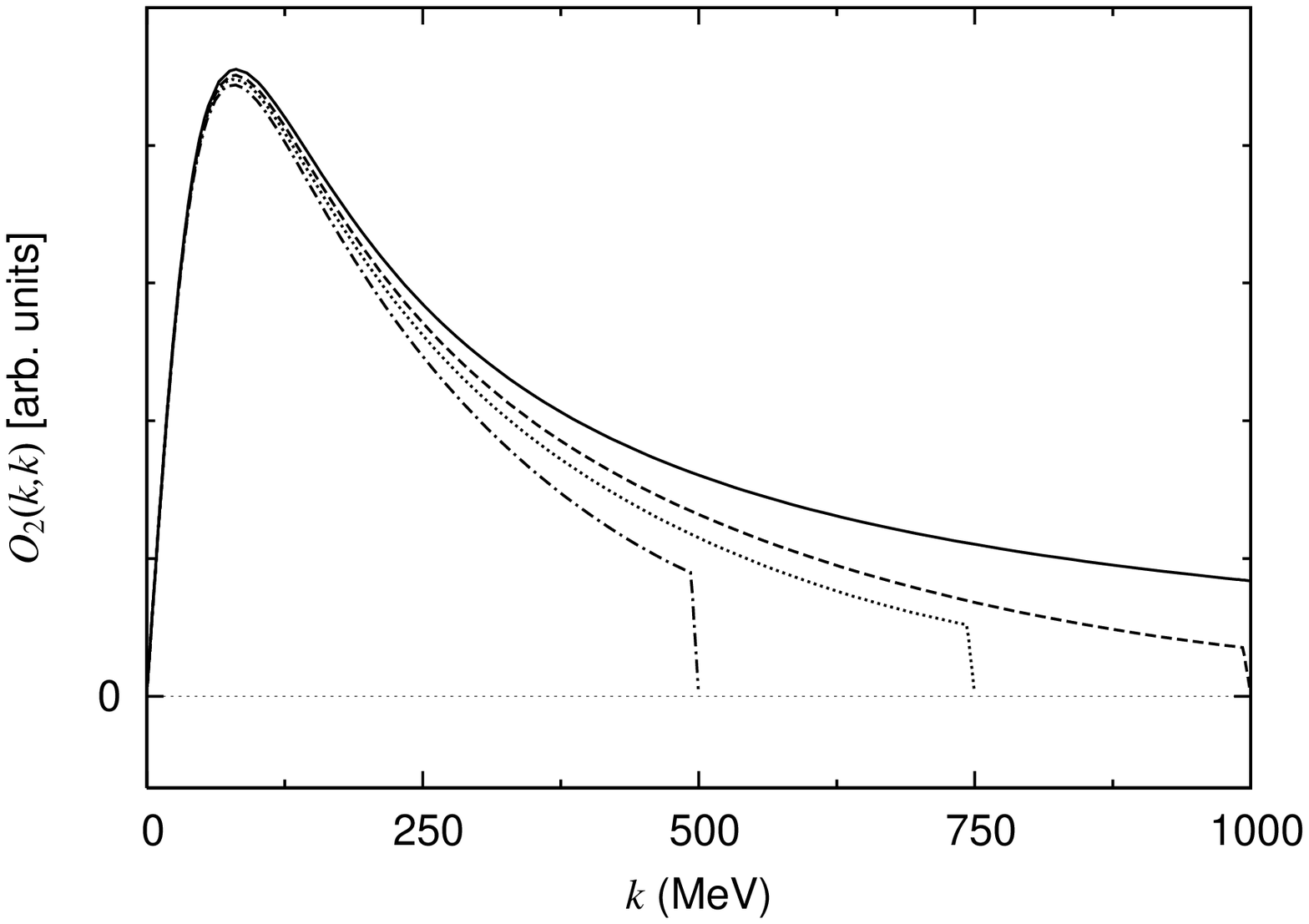}
\caption{\label{fig_run}
Running of the $\pi$ production operator for the $^3P_1\to{}^3S_1$ transition 
in $NN\to d\pi$ ($\eta = 0.1$).
The starting chiral NLO operator is shown by the solid line.
After the RG running, we obtain low-momentum operators for
$\Lambda$~=~1000~MeV (dashed line), $\Lambda$~=~750~MeV (dotted line),
$\Lambda$~=~500~MeV (dash-dotted line).
}
\end{center}
\end{minipage}
\hspace{2mm}
\begin{minipage}[t]{65mm}
\begin{center}
\includegraphics[width=65mm]{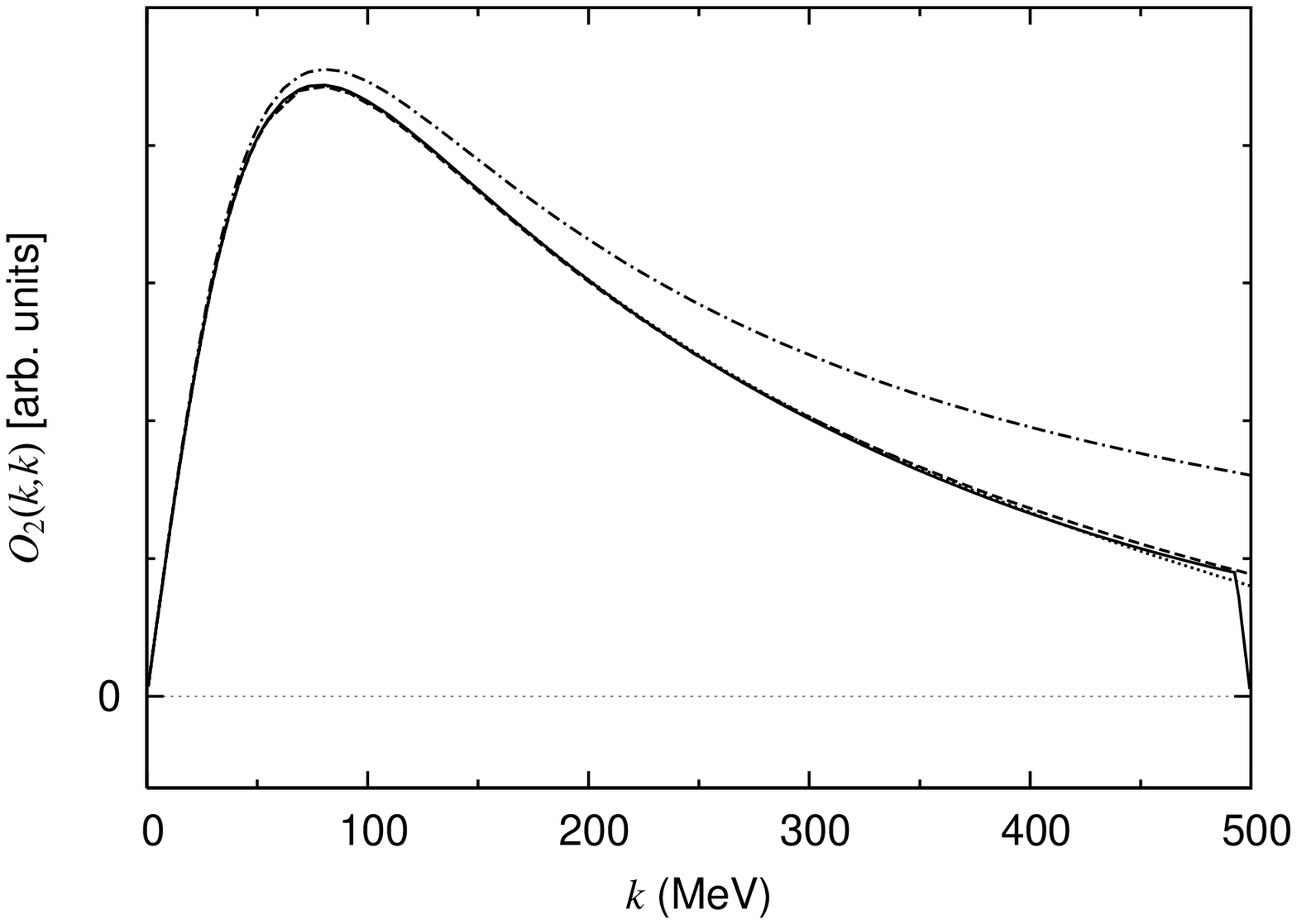}
\caption{\label{fig_diag}
The RG low-momentum pion production operator
($\Lambda$ = 500 MeV, $\eta$ = 0.1), shown by the solid line,
is simulated by the
original NLO operator (dash-dotted line) 
plus a contact term with one nucleon derivative (dashed line).
The dotted line has additional contact terms
with three derivatives.}
\end{center}
\end{minipage}
\end{figure}

The running of the radial part of the $^3P_1\to {}^3S_1$ transition operator 
(diagonal matrix elements) is shown in Fig.~\ref{fig_run}.
The solid line is the starting NLO chiral operator (before 
renormalization). 
After the RG running, we obtain the RG low-momentum operator shown by the 
dashed line ($\Lambda$~=~1000~MeV),
dotted line ($\Lambda$~=~750~MeV), and
dash-dotted line ($\Lambda$~=~500~MeV).
We parameterize this low-momentum operator using the NLO operator plus
the counter terms, omitting the kink part when fitting the counter terms.
In Fig.~\ref{fig_diag} we show the simulation of the low-momentum
operator using the NLO operator plus the counter terms.
The dash-dotted line is the starting NLO chiral operator.
The dashed line contains the NLO operator plus the lowest counter term.
The dotted line contains the additional terms of one higher order.
The renormalized part (the difference between the solid and dash-dotted lines) 
is thus accurately captured by the higher order counter terms of natural
strength.
As for the off-diagonal matrix elements,
we find a similar trend in the simulation using the counter terms.
Our result allows us to conclude that, to the level of precision (and order)
we are working, the $NN\to d\pi$ calculation is renormalized without 
including a chiral-symmetry-violating counter term.
However, this outcome does not necessarily mean that the (sharp) cutoff
regularization does not generate {\it any} chiral-symmetry-violating
operator.
Nevertheless, we could still claim that there is no \emph{practical} ground for
introducing a chiral-symmetry-violating operator for this process.

\section{Summary}\label{sec_summary}
We have studied what happens to the pion production operator when a
cutoff is introduced.
In particular, we are interested in the possibility that the running with 
cutoff requires us to introduce a chiral-symmetry-violating interaction.
We started with the chiral NLO operator~\cite{lensky} for $s$-wave pion 
production in $NN\to d\pi$ and introduced a cutoff by using the WRG equation.
After the RG running, we parameterized the RG low-momentum operator by a chiral
expansion including higher-order counter terms, which are expected to absorb 
contributions from the high momentum states that were integrated out.
We found that the expansion is indeed accurate, with the LECs of natural
strength.
Therefore, we see no evidence that the running with the cutoff generates 
chiral-symmetry-violating interactions.

\section*{Acknowledgments}
This work was supported by the Natural Sciences and Engineering Research 
Council of Canada and the U. S. National Science Foundation through
grant PHY-0457014.

%\begin{thebibliography}{000} %for 3 digits
%\begin{thebibliography}{00}  %for 2 digits

%%%%%%%%%%%%%%%   Author and Subject Index
%\printindex{author}{Author Index}
%\blankpage
%
%\printindex{subject}{Subject Index}
% \blankpage

\begin{thebibliography}{0}    %for 1 digit

%%journal paper
\bibitem{lensky}
V. Lensky {\it et al.}, {\it Eur.\ Phys.\ J.} {\bf A27}, 37 (2006).

\bibitem{rg1}
S. X. Nakamura, {\it Prog.\ Theor.\ Phys.}\ {\bf 114}, 77 (2005).

\bibitem{NA}
S. X. Nakamura {\it et al.}, {\it Phys.\ Rev.\ C} {\bf 74}, 034004 (2006).

\bibitem{NA2}
S. X. Nakamura {\it et al.}, arXiv:0704.3757.
\end{thebibliography}
\end{document}